\documentclass[12pt]{article}
\usepackage{amsfonts}
\usepackage{amsmath, amssymb}
\usepackage{graphicx}
\usepackage{color}
\usepackage{ulem}
\usepackage{epsfig}

\newcommand{\bea}{\begin{eqnarray}}
\newcommand{\eea}{\end{eqnarray}}

\begin{document}

\title{The turnstile mechanism across the Kuroshio current: analysis of dynamics in altimeter velocity fields}

\author{Carolina Mendoza$^1$, Ana M. Mancho$^1$\footnote{Correspondence to: a.m.mancho@icmat.es}, Marie-H\'elene Rio$^2$\\
$^1$Instituto de Ciencias Matem\'aticas. CSIC-UAM-UC3M-UCM. \\Serrano 121. 28006 Madrid, Spain. \\
$^2$CLS -- Space Oceanography Division, Toulouse, France. }





%



%

\maketitle

\begin{abstract}
In this article we explore the ability of dynamical systems  tools to describe transport 
in oceanic flows characterized by data sets measured from satellite. 
In particular we have  studied the  geometrical skeleton describing transport in the Kuroshio region. 
For this purpose we have computed special hyperbolic trajectories, recognized as distinguished hyperbolic trajectories,
that act as organizing centres of the flow.  We have computed their 
stable and unstable manifolds, and they reveal that the turnstile mechanism is at work
during several {spring}  months in the year 2003 across the Kuroshio current. We have found that near the hyperbolic 
trajectories takes place a filamentous  transport front-cross the current that mixes waters at both sides.
\end{abstract}

\section{Introduction}
Transport  processes across the ocean have an important impact in global climate \cite{bowen, dabiri},  as oceans are important reservoirs
of heat and carbon dioxide which are the most important driving forces of climate change. 
Understanding and describing  transport processes on the ocean have been a challenging task over the past years, since
great currents over the ocean such as Gulf stream or Kuroshio that seem  smooth, river-like streams  in  their Eulerian description,
present a messy pattern  on the  float tracking  \cite{bowen, lavender}. Understanding such patterns is an essential 
ingredient for predicting the effects of global warming in Earth's future climate.
 
 Dynamical systems theory has provided in recent years a framework for describing transport
  in time dependent flows. It has been useful for finding transport routes in complex flows, which means it has successfully  found order structures from erratic patterns. 
  Typical tools in dynamical systems theory are Lyapunov exponents and invariant manifolds.
These concepts are defined strictly on infinite time systems, however realistic flows, like those arising in geophysics or
oceanography, are not infinite time systems and for their description,
finite time versions of the definition of Lyapunov exponents, such as,  finite size Lyapunov exponents FSLE \cite{vulpiani} 
and finite time Lyapunov exponents FTLE \cite{nese,haller} are used.  Invariant manifolds are objects that in the classical theory of dynamical systems 
are asymptotically defined for special trajectories such as fixed points or periodic orbits with hyperbolic stability.
Recent articles by \cite{ide,ju,chaos} have provided definitions that generalise  these reference orbits for time dependent aperiodic flows.
These special trajectories are called distinguished trajectories. The definition given in \cite{chaos} is proven to  encompass fixed points and periodic orbits as 
particular cases. Separation points (SP) on boundaries such as, detachment
and reattachment points are also particular examples of distinguished hyperbolic trajectories (DHT). The computation of SP has been  also addressed  in \cite{haller2}
however the definition provided in   \cite{chaos} provides a systematic way for finding these special trajectories not only 
on boundaries but also in the interior of the flow.   DHT and their stable and unstable  manifolds are the essential dynamical tools we will apply 
 in this article for describing transport in geophysical flows.
 
 Lagrangian and Eulerian observational datasets have held great 
promise of providing deep insight  on oceanic transport.
Drifters observations  have revealed fluid exchange \cite{bower1,lozier} in the Gulf Stream.
 An important step forwards in the description of exchange of trajectories across the Gulf Stream 
 was the meandering kinematic model proposed by \cite{bower2}.
 Studies such as  \cite{lozier} have succeeded in describing transport in  simplified
 models of the Gulf Stream by means of lobe dynamics \cite{bower2}. 
 Recently, ocean-observing satellites are entering a new operational era 
 and naturally occupy an important place in programs designed to manage and predict ocean and climate change.
These instruments may provide a near real time processing system of satellite  altimeter data.
 In this article we describe cross-frontal  transport along a major ocean current, the Kuroshio current,
based on a Lagrangian description of  observed  altimeter velocities, over a period of the year 2003. 
Our description uses  dynamical objects such as
the  stable and unstable manifolds of distinguished trajectories.
 We detect the turnstile mechanism across a part of the Kuroshio current. The turnstile
survives persistently from April 3 to  May 26, 2003. This mechanism had already been 
found to be present in realistic numerically produced  data sets modeling the North Balearic front \cite{jpo}. In this article we discuss 
similar ideas  with the challenge of dealing with altimeter data. Lagrangian Coherent Structures (LCS), as the use of finite time Lyapunov exponents 
to compute transport issues in geophysical flows is referred to,
or FSLE have also recently applied to describe dynamics in altimeter datasets 
\cite{beron,rossi,rypina}, although these studies are focused in  areas of the ocean surface different from the Kuroshio current.
 
The structure of the article is as follows. In Section 2 we describe the analysed  data and to what extent the velocity field satisfies  the divergence free approach.
Section 3 describes the details of the interpolation used in the data and  provides the equations of the  continuous dynamical system used on our analysis. 
Section 4 describes the computation of the distinguished hyperbolic trajectories and of their stable and unstable manifolds for the system built up in Section 3. 
Section 5 describes the turnstile mechanism governing the dynamics  across a current that persist on the data amid latitudes $32^o$N and $40^o$N  and longitudes
$155^o$E and    $165^o$E during the months of April to May 2003.  Finally Section 6 details the conclusions.

\section{The altimeter dataset}

\label{sec:altimetry}

In this work we analyze transport across the Kuroshio current from April   3 to
May 26,  2003,   {spanning} a time interval of { 54} days.  

We use daily maps of surface currents computed at CLS in the framework of the SURCOUF project \cite{larnicol}
from a combination of altimetric sea surface heights and windstress data in a two-steps procedure:
On one hand, multimission (ERS-ENVISAT, TOPEX-JASON) altimetric maps of sea level anomaly (SLA) are 
added to the RIO05 global Mean Dynamic Topography \cite{Rio2,Rio4} to obtain global maps 
of sea surface heights from which surface geostrophic velocities ($u_g$, $v_g$) are obtained by simple derivation. 			
\begin{eqnarray}
u_g=-\frac{g}{f}\frac{\partial h}{\partial y} \\
v_g=\frac{g}{f}\frac{\partial h}{\partial x}
\end{eqnarray}
where $f$ is the coriolis parameter and g the gravitational constant.

On the other hand, the Ekman component of the ocean surface current ($u_{ek}$, $v_{ek}$) is estimated using a 2-parameter  model:  
$u_{ek}=be^{i \theta}\tau$ where $b$ and $\theta$ are estimated by latitudinal bands from a least square 
fit between ECMWF 6-hourly windstress analysis $\tau$  
and an estimate of the Ekman current obtained removing the altimetric  geostrophic current from the total current 
measured by drifting buoy velocities available from 1993 to 2005 (the method is described in further details in \cite{Rio3}).

Both the geostrophic and the Ekman component of the ocean surface current are added to 
obtain estimates of the ocean total surface current that are used in the Kuroshio area for the present study.

\begin{figure*}[t]
\vspace*{2mm}
\begin{center}
\includegraphics[width=14cm]{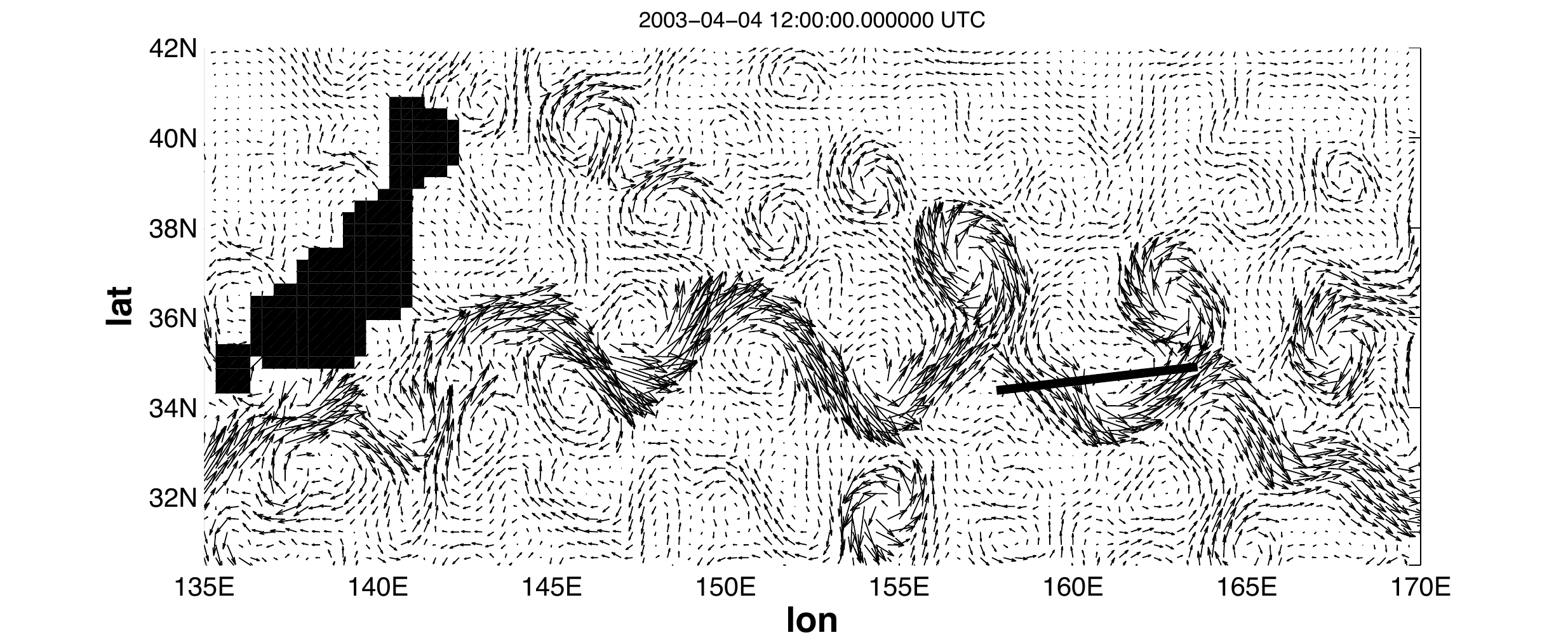}
\end{center}
\caption{Velocity field of the Kuroshio current on April 4, 2003. A line marks a piece of current across which 
transport is studied. Maximum
values of the velocity field are about { 3.65 m/s}. }
\label{velocityfield}
\end{figure*}

Fig. \ref{velocityfield} shows a sample of the velocity field  on April 4,  2003.
We have marked with a line the current across which we  study 
transport. The presence in the nearby of two counterclockwise eddies 
 suggest the presence of two hyperbolic points at both sides of the line
 that could maintain the turnstile mechanism.  We will show in Section 5 that this is truly case.
 The interaction of these type of eddies with the main current has been pointed out 
 to trigger the Kuroshio meander formation \cite{waseda}.

Our work considers that in the area under study the particle motion is restricted to the ocean surface, i.e,  there is no significative vertical motions. 
The altimeter-derived velocity field is two-dimensional, however since the Ekman correction 
has been introduced, it is not guaranteed that the data is divergence free.  We have  computed the divergence of the 
velocity field   in the whole { period }  and  area under 
study and we have verified  that despite these corrections  the divergence free approach is still valid.

\section{The equations of motion}
The equations of motion that describe the horizontal evolution of
particle trajectories on a sphere is
\begin{eqnarray}
\frac{d \phi}{d t}&=&\frac{u(\phi,\lambda,t)}{R {\rm cos}(\lambda)}\label{eqm1}\\
\frac{d \lambda}{d t}&=&\frac{v(\phi,\lambda,t)}{R}\label{eqm2}
\end{eqnarray}
where $u$ and $v$ represent the eastward and northward
components of the altimetry surface velocity field { respectively} described in the previous section. 
Particle trajectories must be integrated
in equations (\ref{eqm1})-(\ref{eqm2}) and since information is provided just in a
discrete space-time grid, a first issue to deal with is that of
interpolation of discrete data sets. A recent paper by \cite{msw}
compares different interpolation techniques in tracking particle
trajectories. Bicubic spatial interpolation in space \cite{nr}
and third order Lagrange polynomials in time are shown to provide
a computationally efficient and accurate method. We use this
technique in our computations. 
However we notice that bicubic
spatial interpolation in space as discussed in \cite{nr} requires
an equally spaced grid. Our data input is expressed in spherical
coordinates, and the grid is not uniformly spaced in the latitude
coordinate. In order to interpolate in an uniformly spaced grid,
we transform our coordinate system ($\lambda, \phi$) to a new
one ($\mu, \phi$). The latitude
$\lambda$ is related to the new coordinate $\mu$ by
\begin{equation}
\mu=   {\rm ln}|{\rm sec}\lambda+ {\rm tan}\lambda| \label{mul}
\end{equation}
Our velocity field is now on a uniform grid in the ($\mu, \phi$)
coordinates. The equations of motion in the old variables are
transformed to a new expression in the new variables,
\begin{eqnarray}
\frac{d \phi}{d t}&=&\frac{u(\phi,\mu,t)}{R \, {\rm cos}(\lambda(\mu))} \label{sd1}\\
\frac{d \mu}{d t}&=&\frac{v(\phi,\mu,t)}{R \,{\rm
cos}(\lambda(\mu))} \label{sd2}
\end{eqnarray}
where $\lambda(\mu)$ is obtained by inverting Eq. (\ref{mul}),
i.e.
\begin{equation}
\lambda=\frac{\pi}{2}-2 \,{\rm atan}(e^{-\mu}) \ . \label{lmu}
\end{equation}
Once trajectories are integrated from these equations, for
presentation purposes one can convert $\mu$ values back to
latitudes $\lambda$ just by using (\ref{lmu}). 

 \begin{figure*}[t]
\vspace*{2mm}
\begin{center}
a)\includegraphics[width=7cm]{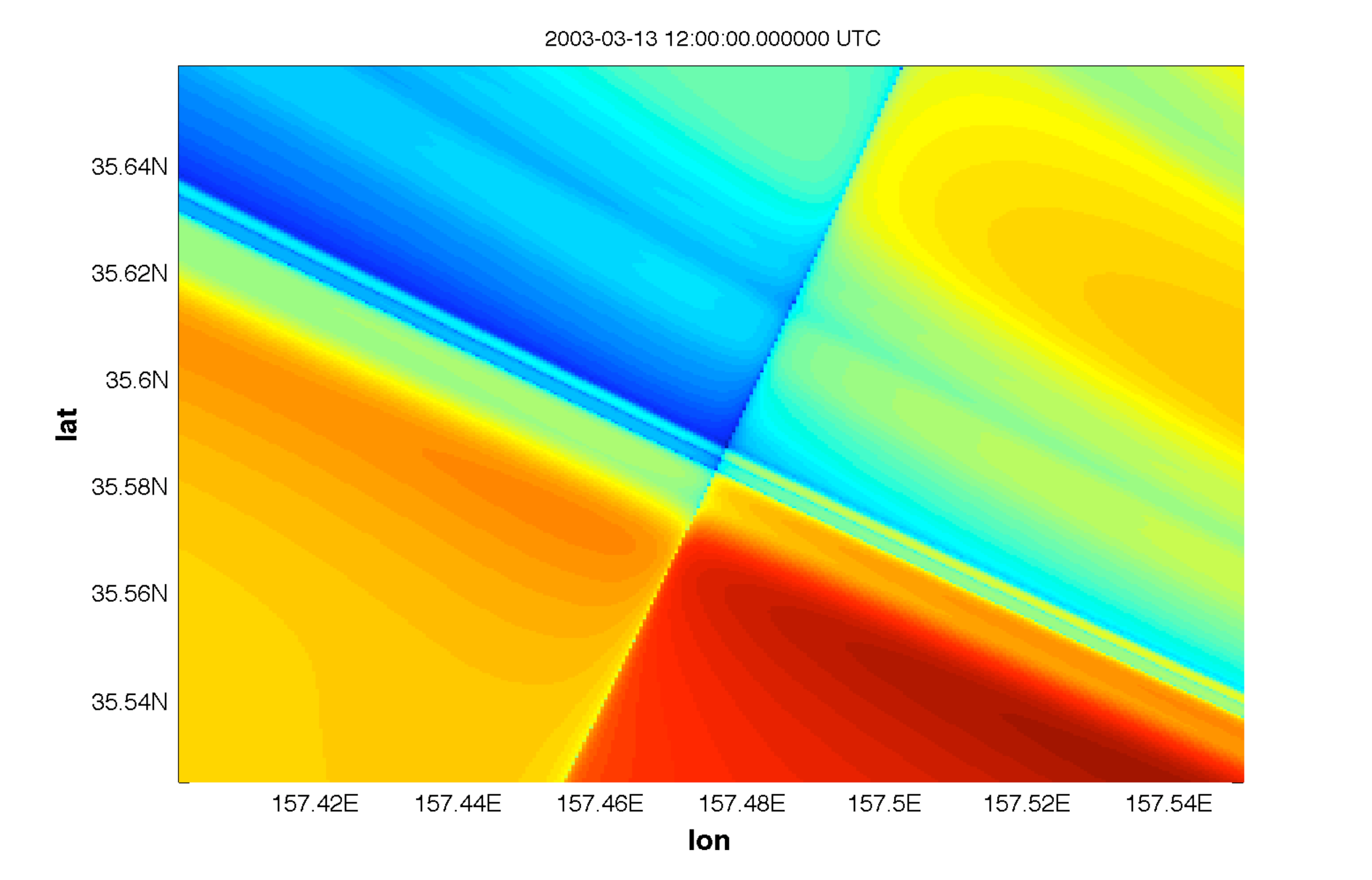}b)\includegraphics[width=7 cm]{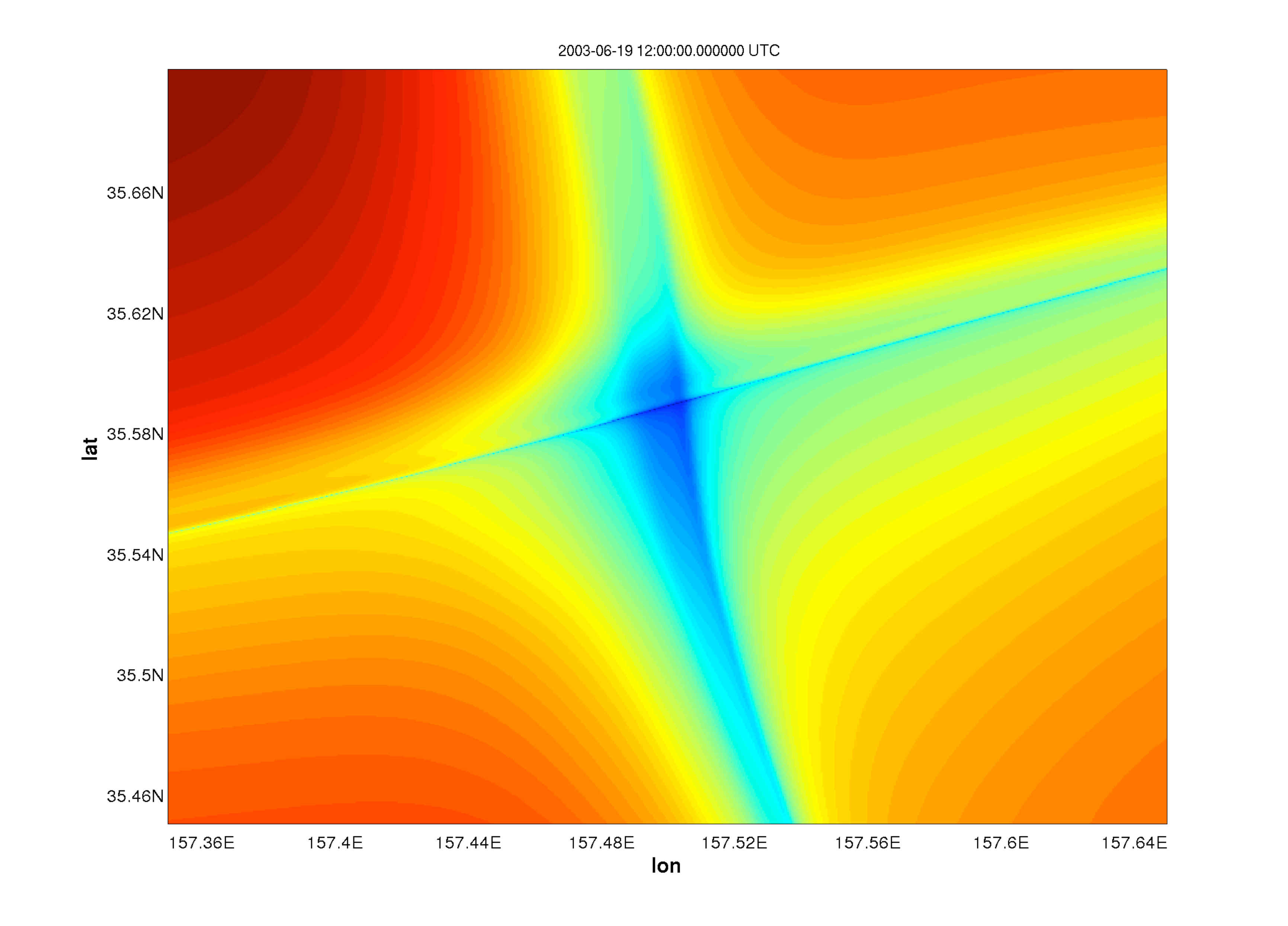}
\end{center}
\caption{ Structure of the function $M$ for $\tau=15$,  a) in the neighbourhood of the western DHT,  DHT$_W$, on March 13, 2003; b)  in the neighbourhood of the eastern DHT,  DHT$_E$, on  June 19, 2003.}
\label{matrizM}
\end{figure*}

\section{Distinguished Hyperbolic Trajectories and Manifolds }

Distinguished hyperbolic trajectories and their unstable and
stable manifolds are the dynamical systems objects used to describe and quantify transport. 

Fig. \ref{velocityfield} marks with a line the eulerian feature across which we study transport.
The current, similarly to that studied in \cite{jpo,cw},  flows eastward. First we identify  a DHT in the western part  
of the flow,  for which we compute the unstable manifold  and a DHT in the eastern part, for which
we compute the stable manifold.

Computation of distinguished hyperbolic trajectories for aperiodic flows has been discussed in \cite{ide,ju,nlpg,chaos}. 
The approach taken in this article is that of  \cite{chaos}, which is based on the function $M$ defined  as follows.
Let $({\phi(t), \mu}(t))$ denote a trajectory of the system (\ref{sd1})-(\ref{sd2}).
For all initial conditions  $({\phi^*, \mu^*})$ in an open  domain ${\mathcal B}$ of the ocean surface,  at a given time $t^*$,
consider the  function $M(\phi^*, \mu^*,t^*)_{\tau}:{\mathcal B \times t}\to\mathbb{R}$ 
\begin{equation}
M= \left(  \int^{t^*+\tau}_{t^*-\tau} \! \!\!\sqrt{\left(\frac{d \phi(t)}{dt}  \right)^2+ \left(\frac{d \mu(t)}{dt} \right)^2 }dt \right), \label{def:M}
\end{equation}
For an initial condition $(\phi^*,\mu^*)$ at $t^*$, the function $M$ measures the length of the curve outlined 
by a trajectory from $t^*-\tau$ to  $t^*+\tau$ on the  plane $(\phi,\mu)$.  

We discuss in more detail the numerical evaluation of
$M$ as defined in Eq. (\ref{def:M}).  Trajectories $({\phi(t), \mu}(t))$ of the system (\ref{sd1})-(\ref{sd2}) are obtained numerically,
and thus represented by  a finite number of points, $L$. A discrete version of Eq. (\ref{def:M}) is:
\begin{equation}
M= \sum_{j=1}^{L-1} \left(  \int^{p_f}_{p_i} \! \!\!\sqrt{\left(\frac{d \phi_j(p)}{dp}\right)^2+\left(\frac{d \mu_j(p)}{dp}\right)^2 }dp \right), 
\end{equation}
where the functions $\phi_j(p)$ and $\mu_j(p)$ represent a curve interpolation parametrized by $p$, and the integral
\begin{equation}
\int^{p_f}_{p_i} \! \!\!\sqrt{\left(\frac{d \phi_j(p)}{dp}\right)^2+\left(\frac{d \mu_j(p)}{dp}\right)^2 }dp \label{integral}
\end{equation}
is computed  numerically. Following the methodology described in \cite{chaos} we have used 
 the interpolation method used by  \cite{dr} in the context of contour dynamics. To compute the integral (\ref{integral})
we have used the Romberges method (see \cite{nr}) of  order $2K$ with $K=5$. 

The function $M$ is defined over an open set, so it does not necessarily attain a 
minimum, but if it does, the minimum is denoted by  ${\rm min}(M_{\tau})$. In \cite{chaos} it is found that the position of the 
minimum of $M$ at each $t^*$ is a function of $\tau$.  For   $\tau>>0$, the minimum  converges to a fixed value  called {\it limit coordinate}. Distinguished trajectories
are those that for a time interval, pass close enough (at a distance $\epsilon$, typically within numerical accuracy) to a path of limit coordinates. Computing
 limit coordinates becomes a method for finding distinguished  trajectories because it is quite usual that there exist trajectories passing near these paths.
 
\begin{figure*}[t]
\vspace*{2mm}
\begin{center}
a)\includegraphics[width=7cm]{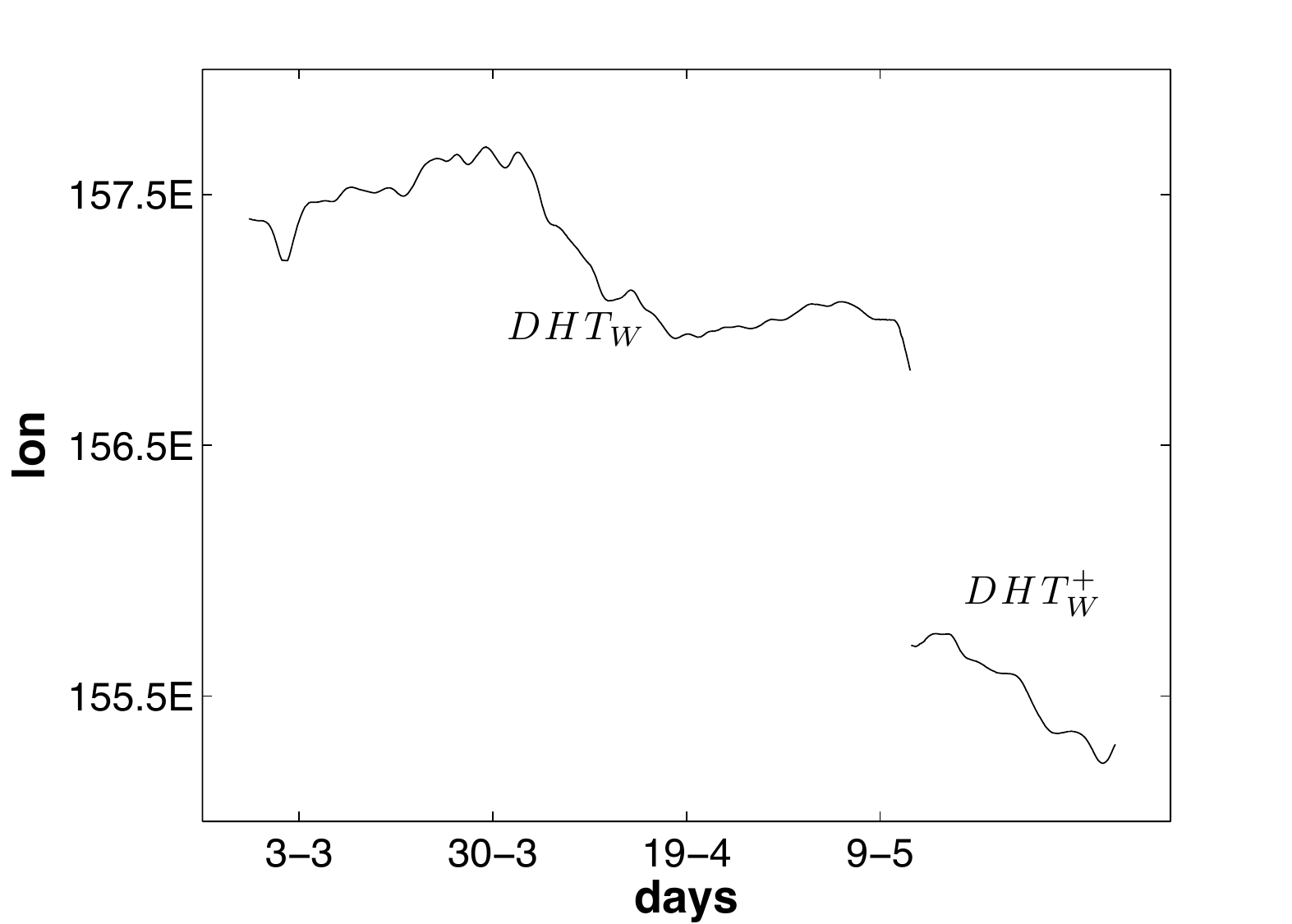}b)\includegraphics[width=7 cm]{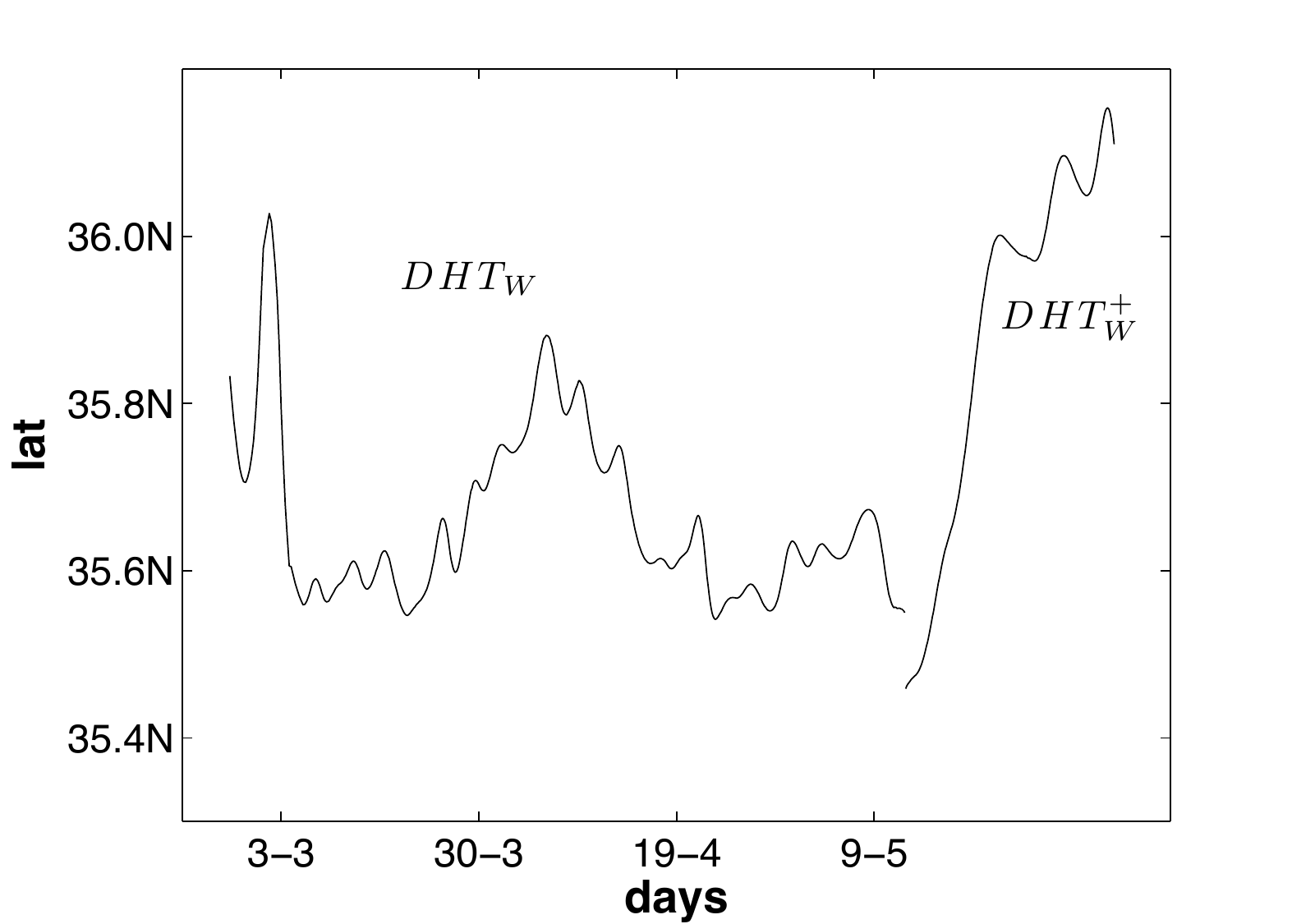}
\end{center}
\caption{ Path of limit coordinates for DHT$_W$ and DHT$^+_W$; a) longitude vs. time; b) latitude vs. time.}
\label{dhtl2}
\end{figure*}

\begin{figure*}[t]
\vspace*{2mm}
\begin{center}
a)\includegraphics[width=7cm]{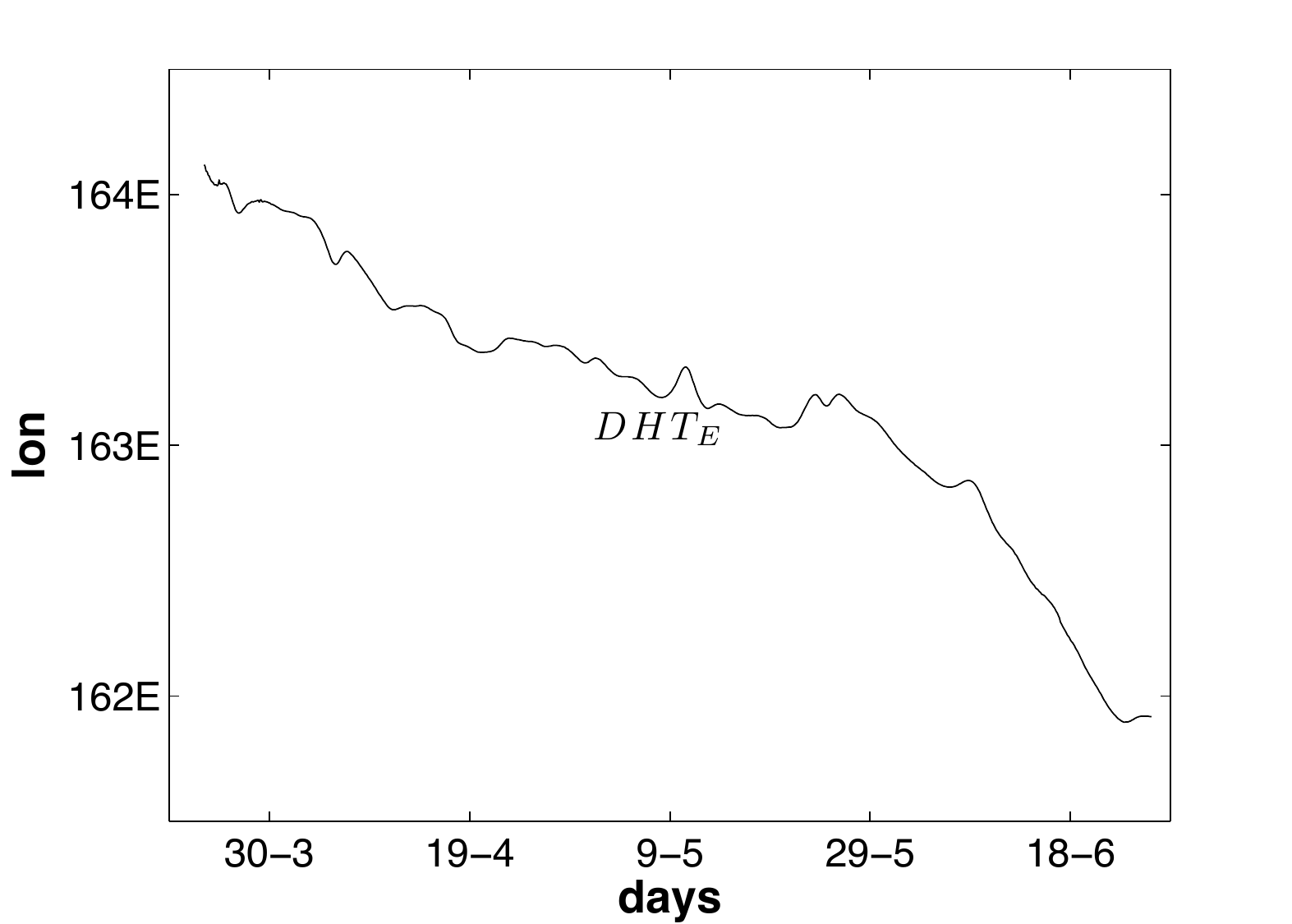}b)\includegraphics[width=7 cm]{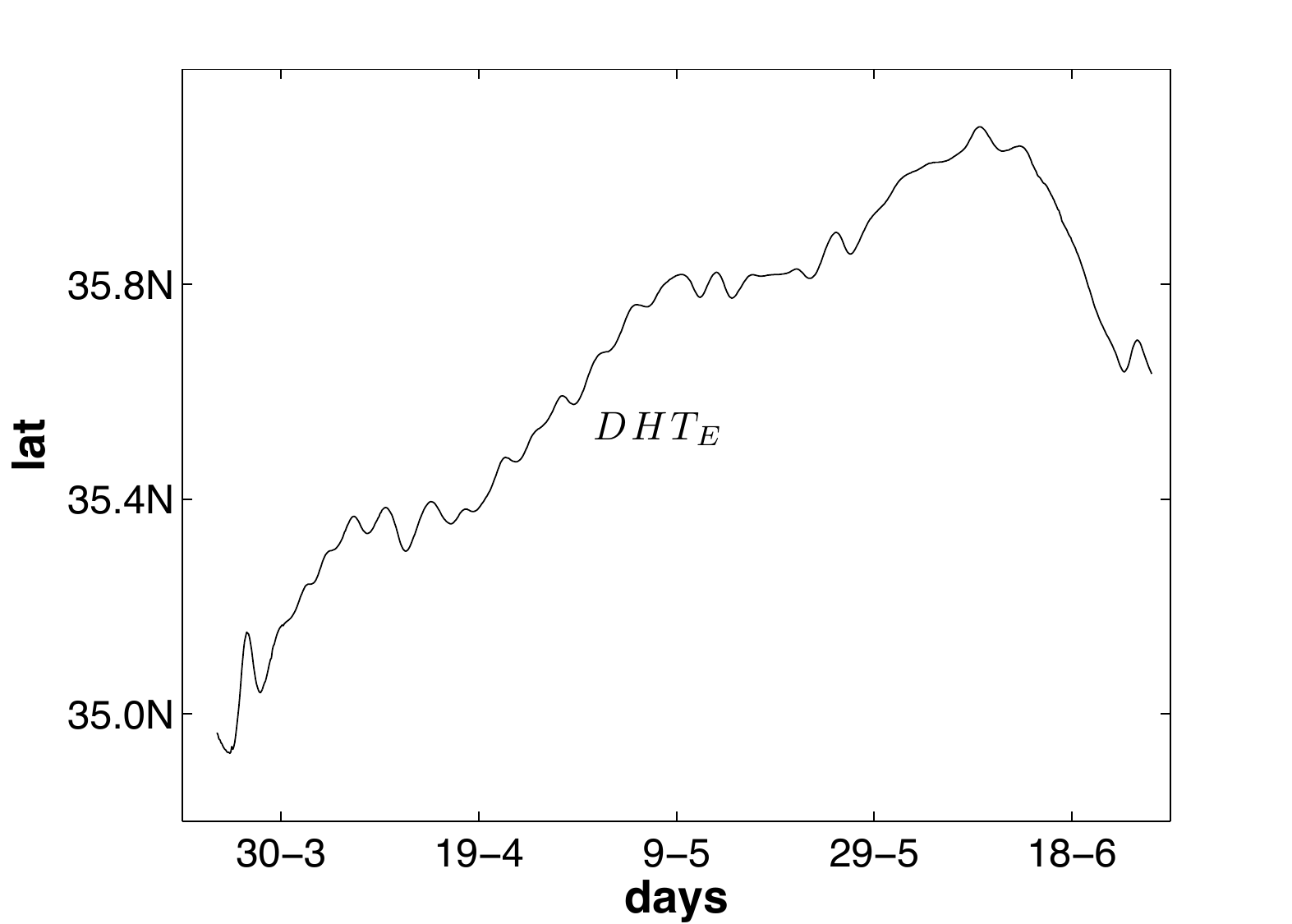}
\end{center}
\caption{ Path of limit coordinates for DHT$_E$; a) longitude vs. time; b) latitude vs. time.}
\label{dhtl3}
\end{figure*}

We have computed  limit coordinates at  opposite sides of the line depicted in Fig. \ref{velocityfield}.  
A typical structure of the function $M$ for $\tau=15$ appears in Fig. \ref{matrizM}.  The color code reads as follows: maximum $M$ values are dark red, 
while minima are dark blue. As discussed in \cite{prlg} the important information provided by $M$ are its minima and its singular features, so no
specific colorbar is required.
Fig.  \ref{matrizM}a) shows the contour plot near the western DHT, DHT$_W$,  in {March 13, 2003}; Fig.  \ref{matrizM}b) depicts the
contour plot of $M$ for the eastern DHT,  DHT$_E$, in { June 19, 2003}. 
The time evolution 
of two paths of limit coordinates, DHT$_W$ and DHT$^+_W$, at the West are displayed in Fig. \ref{dhtl2}.
  Initial conditions on each limit path evolve staying near it for a certain 
time interval. However  
to prove that a trajectory stays close to the path in  long time intervals it is a difficult task because it is not possible to obtain the whole DHT by direct integration methods. 
As  discussed in \cite{chaos},  DHTs are  elusive and any initial condition  in their neighborhood, eventually  leaves it through the unstable manifold. 
If paths in  Fig. \ref{dhtl2} are trajectories they should satisfy Eqs. (\ref{sd1})-(\ref{sd2}). However verifying this requires the computation of the time derivative of the paths 
of limit coordinates and these are very inaccurate.
A better choice  for confirming their trajectory character is to verify that they are at a distance $\epsilon$ of the intersection
of the stable and unstable manifolds. The trajectory DHT$_W$ remains distinguished from  March 5 to May 11, 2003. Outside this time interval the path of limit coordinates although still exists 
 for a short period,  it is not longer a trajectory and eventually the path is lost. This kind of behavior coincides with that described in \cite{chaos} for a similar DHT in a highly aperiodic flow.
At the western end there exists a second DHT  labelled as DHT$^+_W$, which  remains distinguished in an approximately complementary time interval, between  May 10 and June 1, 2003. 
The  distinguished property  of DHT$^+_W$ ends similarly to DHT$_W$. 
 Panels in Fig. \ref{dhtl3} show the path of limit coordinates at the East.  The trajectory DHT$_E$  remains  distinguished between  March 25 and June 24, 2003. 
 
 The fact that paths of limit coordinates are close to trajectories, at least for a certain interval of time, 
indicate that the minima of $M$ provide {\it  Lagrangian} information.
 When the motion described in Eqs. (\ref{sd1})-(\ref{sd2}) is  expressed in a new reference  frame,  trajectories  are related to the  ones in the old frame 
  by the same coordinate transformation. A question then is:  do   trajectories hold in the new frame the distinguished property?. The answer is yes. 
  An example is reported for instance  in \cite{chaos}, where  limit coordinates for the rotating and non rotating Duffing equation
  are related by the   appropriate coordinate transformation. 

Once appropriate DHTs have been identified we compute its stable and unstable manifolds. 
A novel technique
to compute  stable and unstable manifolds of hyperbolic
trajectories in aperiodic flows is developed by \cite{mswi}. 
In
\cite{nlpg,jpo} this manifold computation is  successfully
applied to quite realistic flows.  We will now apply these same algorithms to compute  stable and unstable manifolds 
of DHT in altimeter data sets.  
In our study we compute the unstable manifold of DHT$_W$,   and the stable manifold of DHT$_E$. 
The manifold computation requires as an input the 
position of the DHT at a given time and also the direction of the unstable or stable subspace at that time. In this work
we obtain this in a different way to that proposed in \cite{nlpg}. We take advantage of
 the structure of the function $M$ depicted in Fig.  \ref{matrizM}, which shows strong sharp features.
 In \cite{prlg}   these lines are reported to be an advected structure of  Eqs. (\ref{sd1})-(\ref{sd2}), thus confirming
they are
aligned with the stable and unstable subspaces. We use these directions as input for  the unstable or stable subspaces in the manifold algorithm.
 
\begin{figure}[t]
\vspace*{2mm}
\begin{center}
\includegraphics[width=9.3cm]{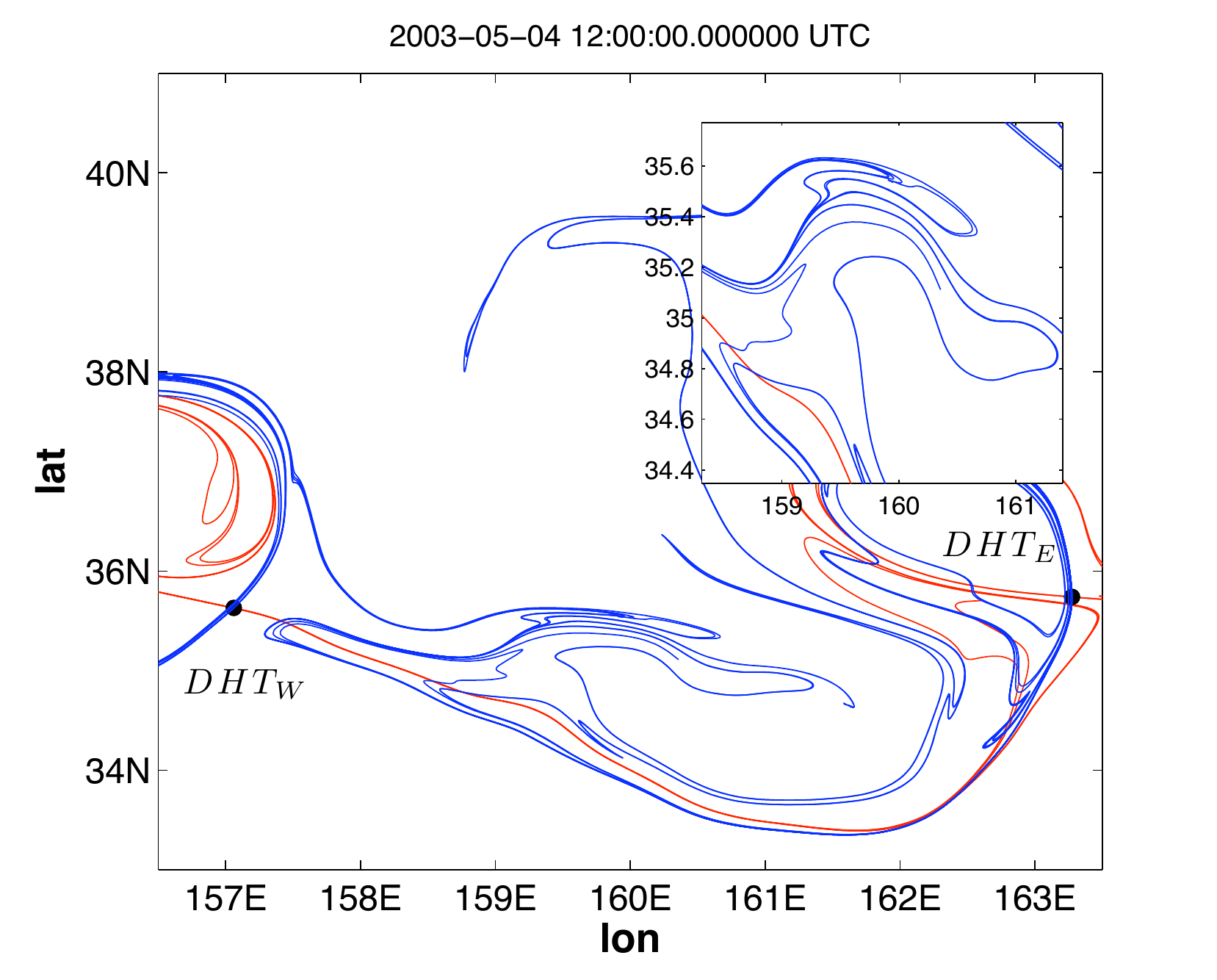}
\end{center}
\caption{Stable (blue) manifold of  DHT$_E$  at the eastern end of the current and unstable (red) manifold of  DHT$_W$ 
 at the west.}
\label{fullmanifolds}
\end{figure}

 Fig. \ref{fullmanifolds} shows the computed manifolds and DHTs on 
 May 4, 2003. Almost  every distinguishable line in this figure
contains numerous foldings of each manifold, thus confirming how intricate they may be.  
In our computations the unstable manifold at a time $t_*$ is made  of trajectories   that at time $t_0$, $t_0<t_*$, were at a small segment aligned with the unstable subspace of the DHT. This is 
a finite time version of   the asymptotic condition
 required for unstable manifolds in infinite time systems. 
 Similarly  the stable manifold at a time $t_*$ is made  of trajectories that at time $t_1$, $t_1>t_*$ are in a small segment aligned with the stable subspace
 of the DHT.
 Other approaches such as  FTLE or FSLE compute  manifolds at a given time  as ridges of a scalar field, 
 providing pieces of curves that are material curves.  However links between pieces of curves are  difficult to establish as they  fade away.
   The  intricate curves of Fig. \ref{fullmanifolds} are linked curves (due to the asymptotic condition imposed in their computation)
    and this is an advantage over other methods.  Manifolds in Fig. \ref{fullmanifolds} however
 are too complex to concrete transport problems so that in next section we will extract from them pieces with  the appropriate  dynamical information.

\section{The turnstile mechanism}

The turnstile mechanism, extensively used and explained in the literature \cite{maw,rlw}, has been found 
to play a role in transport in several oceanographic contexts \cite{jpo,cw}. 
Our manifold computations reveal the presence of the turnstile mechanism transporting masses of water across
the current of Fig. \ref{velocityfield} from north to south and viceversa.  We study it for a period of
54 days from April  3 to  May 26, 2003.

\begin{figure*}[t]
\vspace*{2mm}
\begin{center}
\includegraphics[width=7cm]{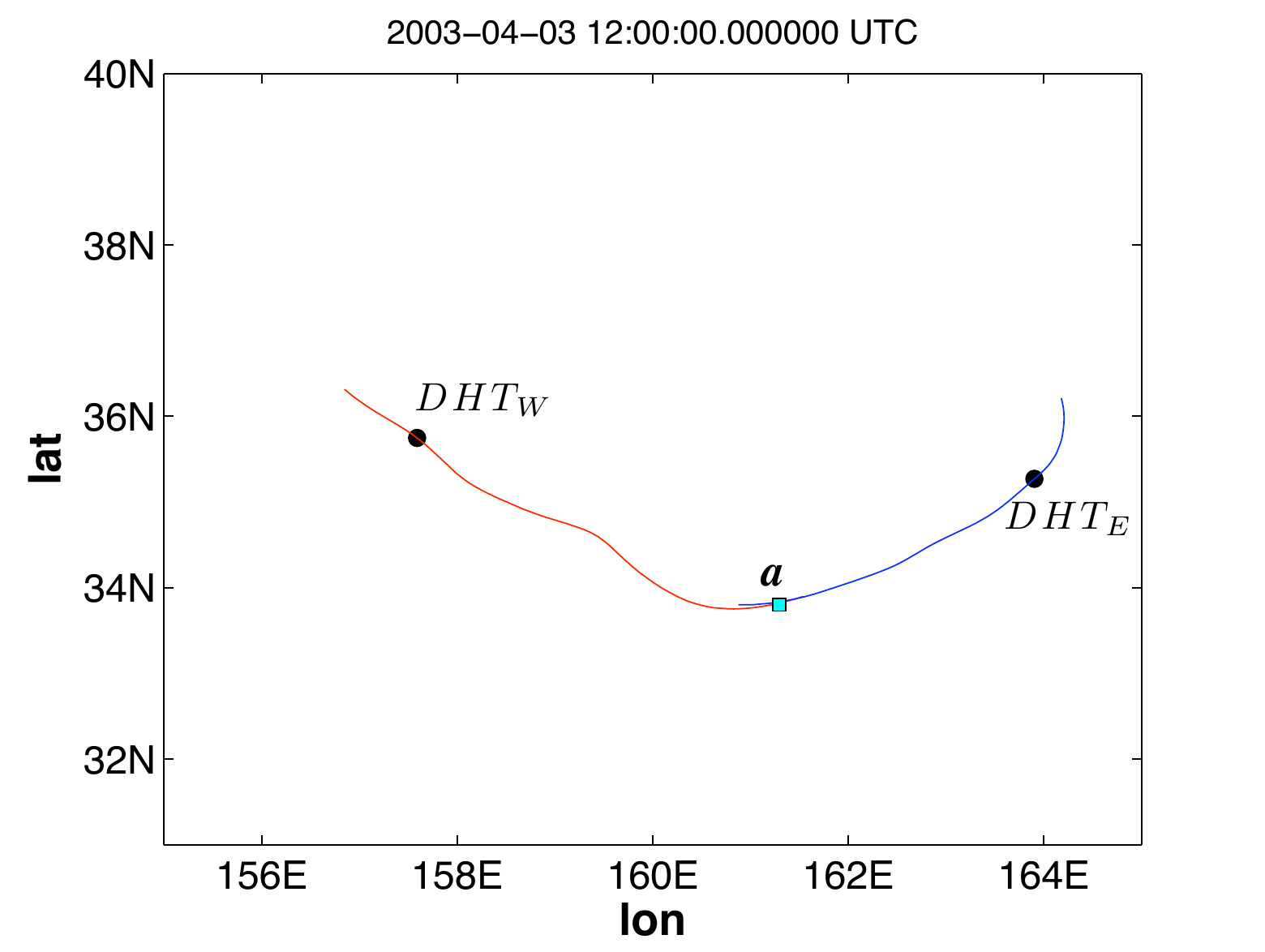}\includegraphics[width=7cm]{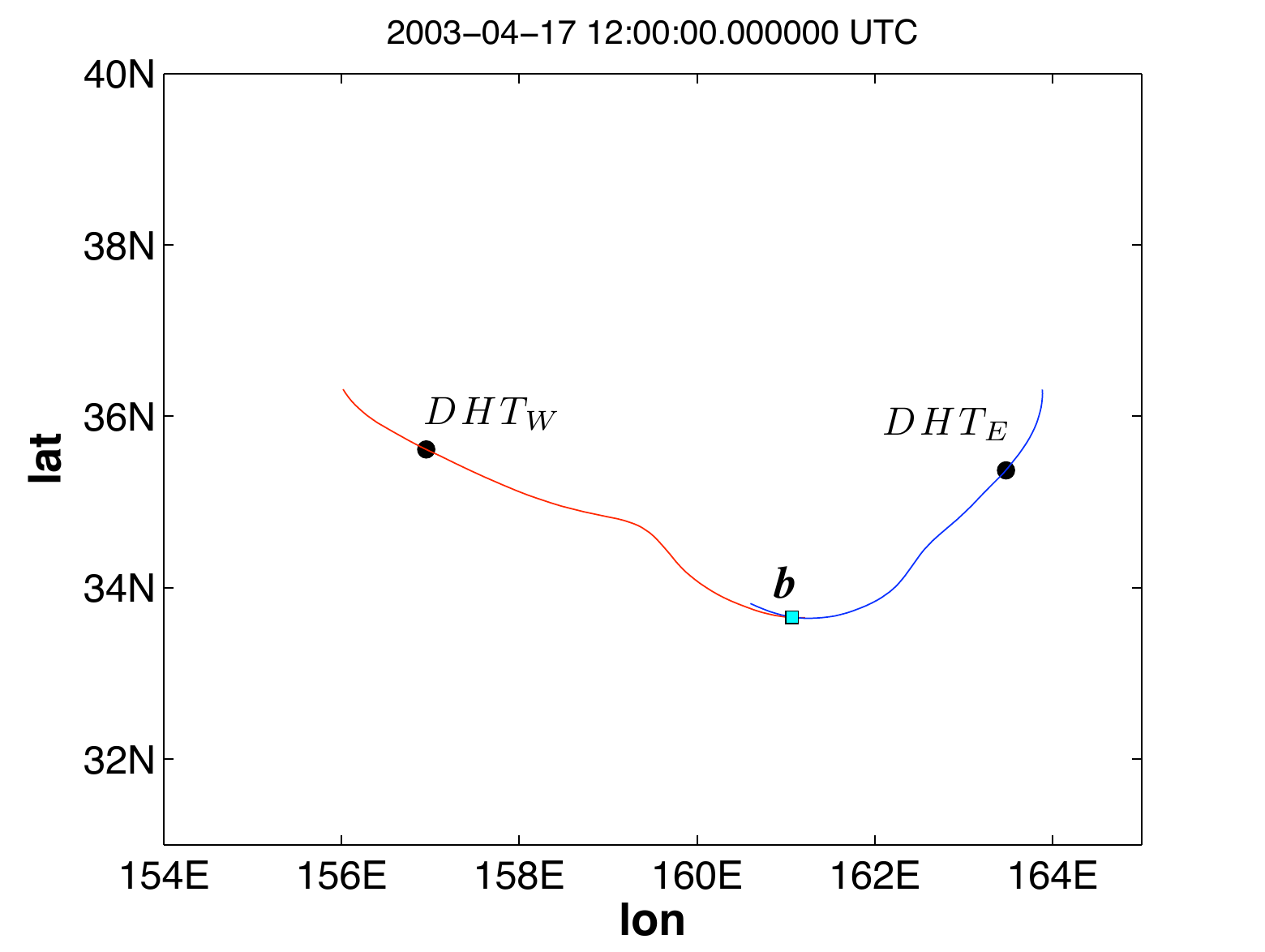}
\end{center}
\caption{Boundaries at days April 3 2003 and April 17 2003 constructed from a (finite length) segment of the
unstable manifold of DHT$_W$ and a (finite length) segment of the stable manifold of DHT$_E$. The boundary intersection points are denoted respectively by $a$ and $b$. }
\label{barrier}
\end{figure*}

 First, we determine a 
time dependent Lagrangian barrier made of pieces of manifolds  that separates north from south. 
Manifolds are made of trajectories, thus as no particle trajectories can cross them, they are barriers to transport.
Fig. \ref{barrier} shows barriers at days {April 4 and April 17}, by
depicting a piece of the unstable manifold of DHT$_W$ and a piece of the stable  manifold of DHT$_E$.
How to choose these portions from intricate curves such as those depicted in Fig. \ref{fullmanifolds}?
First, we consider that a manifold has two branches separated by the DHT, which  constitutes
a reference point. The selections are relatively short  length curves either at one  or both
sides of the  DHT. For instance for the unstable manifold case, these curves are traced up by all the trajectories that expanded from a small segment along
 the unstable subspace of the DHT  during {\it a relatively short  interval of  time}. If  segments along the unstable 
 subspace are  expanded for longer time intervals, then manifolds become so intricate as those in Fig.   \ref{fullmanifolds}.
In Fig. \ref{barrier}   the boundary intersection points are marked
with letters $a$ and $b$.  Intersection points satisfy the property of  invariance, which means that  if the stable and unstable manifolds
intersect at a time in a point, then they intersect for all time, and the intersection point is then a trajectory. 
In order to  help us to understand the time evolution of lobes, we will depict the position of 
trajectories $a$ and $b$ at different times.  

\begin{figure*}[t]
\vspace*{2mm}
\begin{center}
\includegraphics[width=7cm]{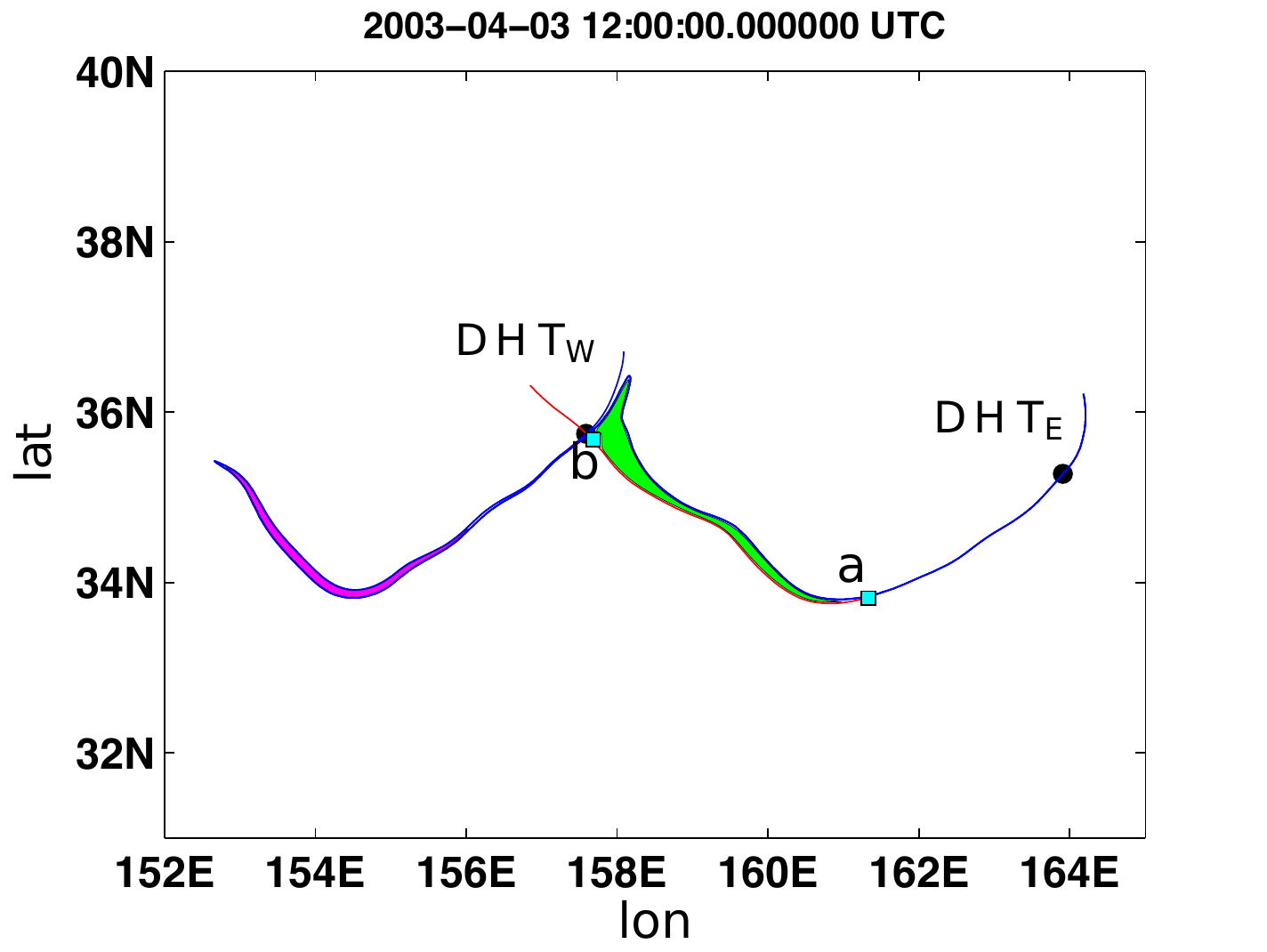}\includegraphics[width=7 cm]{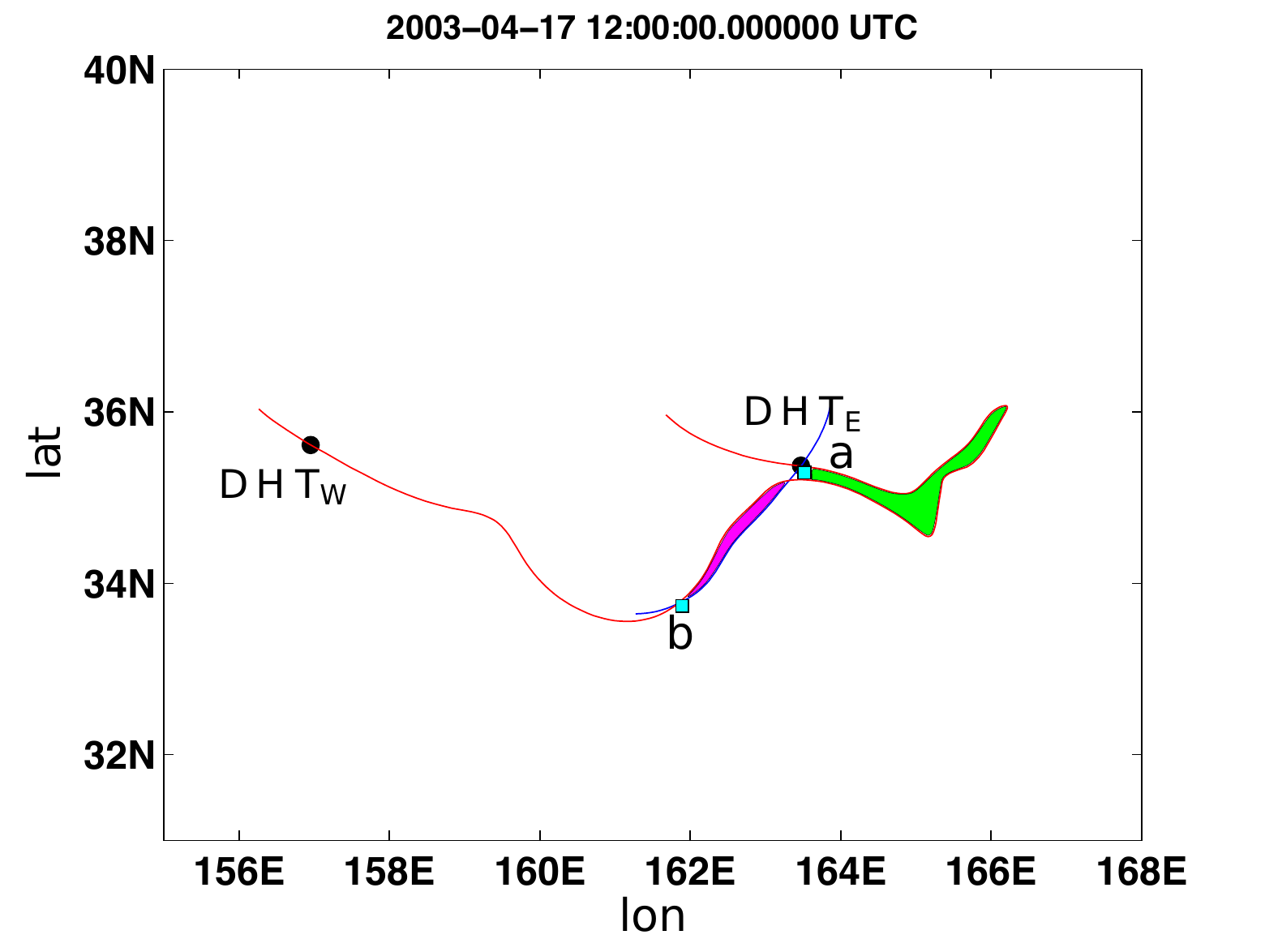}
\end{center}
\caption{ Turnstile lobes at days April 3, 2003 and April 17, 2003. The intersection trajectories $a$ and $b$ are displayed at both days showing their evolution from the DHT$_W$ towards the DHT$_E$. The green area evolves from north to south while the magenta area does from the south to the north.}
\label{turnstile}
\end{figure*}

Fig. \ref{turnstile} shows longer pieces of the unstable and stable manifolds at the same days selected in Fig. \ref{barrier}. 
Manifolds intersect  forming regions called lobes.  It is only the fluid that is inside the lobes that can participate in
turnstile mechanism. Fig. \ref{turnstile} displays two snapshots showing the evolution of  lobes from 
April 2 to April 17. There it is noticed how the lobe which is at the north of the barrier at April 2 crosses
 to the south of the barrier at April 17. Similarly
the lobe which is at the south in April 2, crosses to the north at April 17. Trajectories $a$ and $b$ are depicted showing that 
they evolve approaching to DHT$_E$. The area surrounded by a lobe has been coloured with green if near the western DHT, DHT$_W$, it was at the north side,
and magenta if it was at the south. The turnstile mechanism transports magenta to the north and green to the south.

\begin{figure*}[t]
\vspace*{2mm}
\begin{center}
a)\includegraphics[width=7cm]{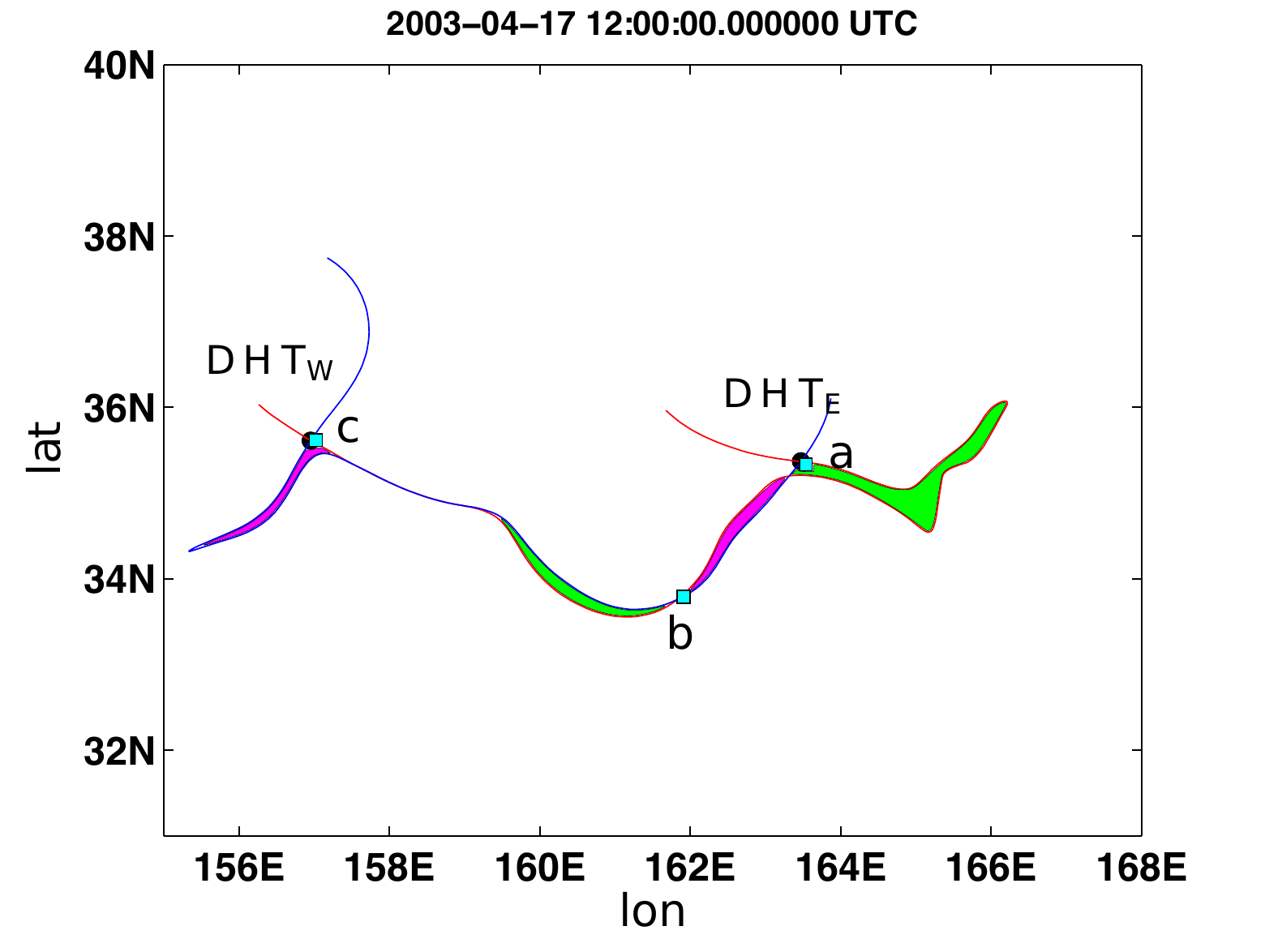}b)\includegraphics[width=7 cm]{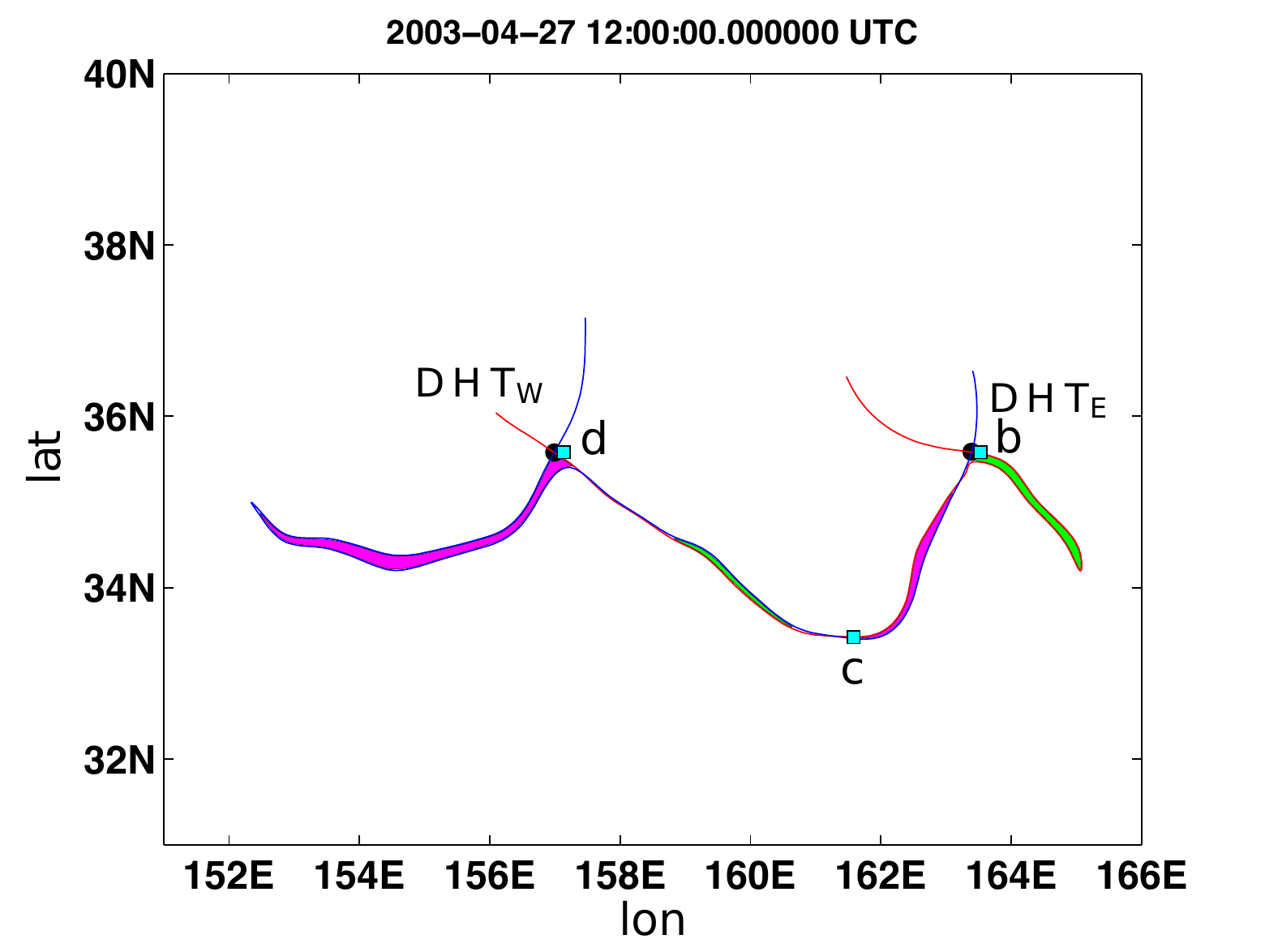}\\
c)\includegraphics[width=7cm]{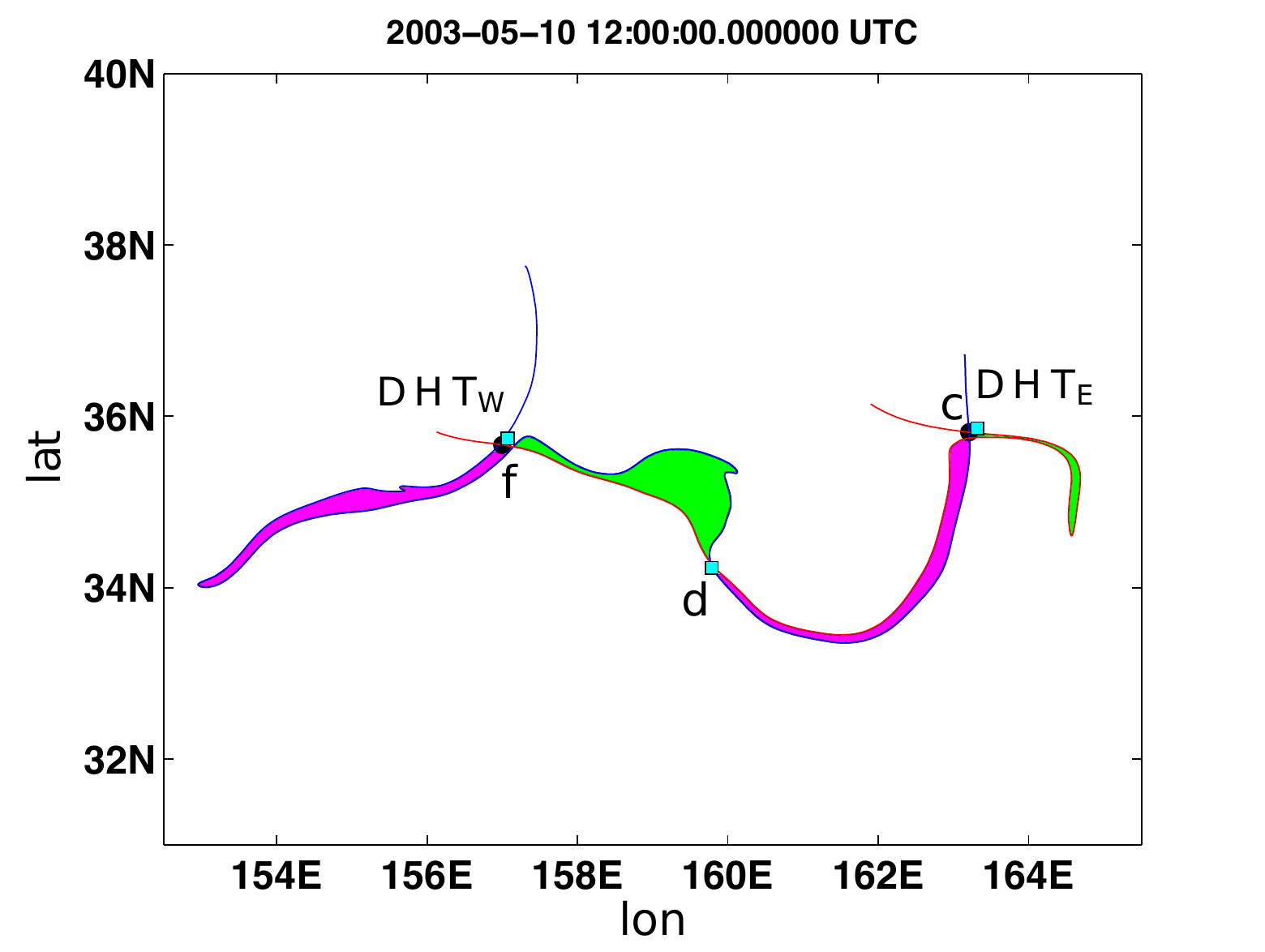}d)\includegraphics[width=7 cm]{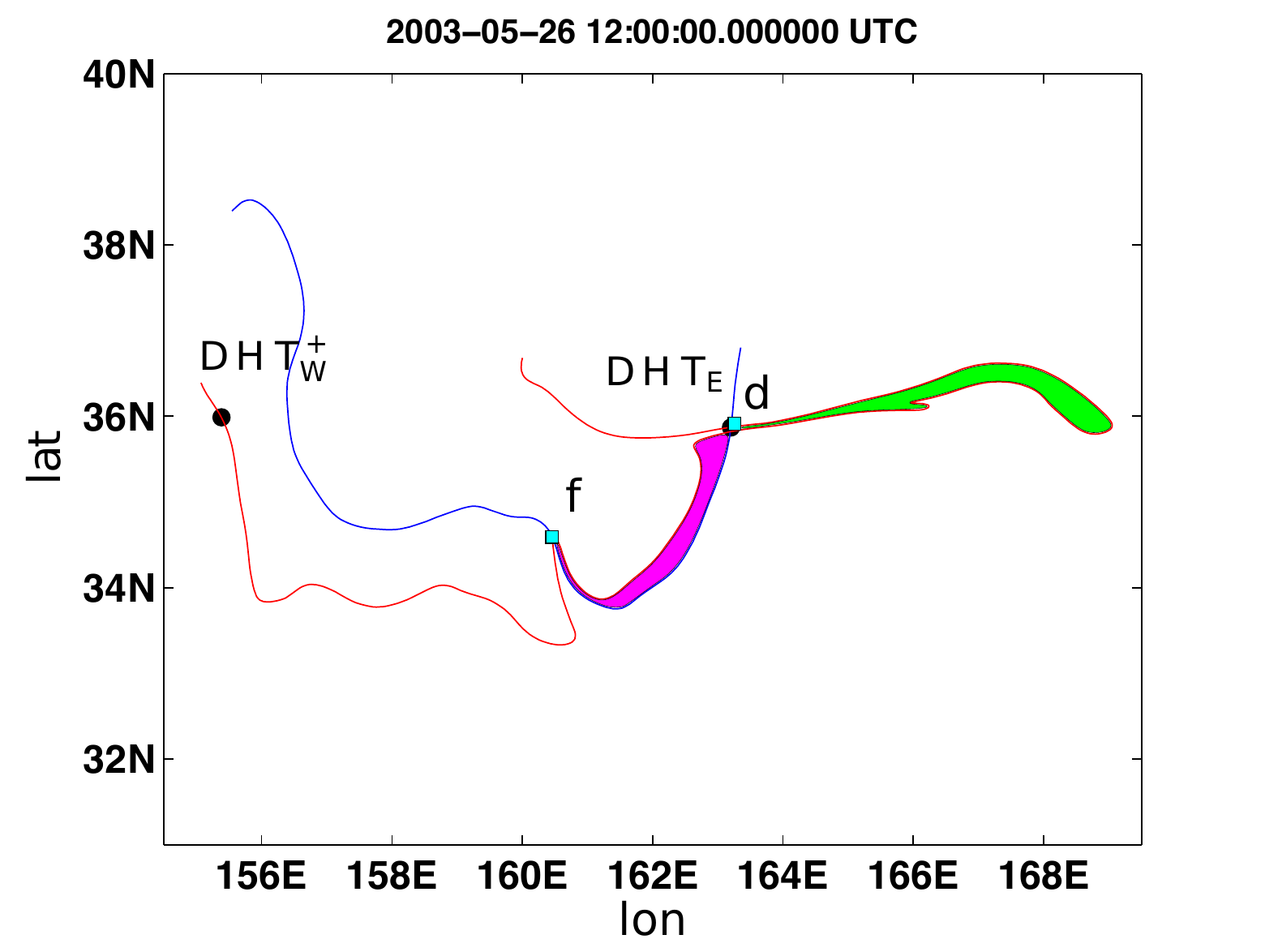}\\
\end{center}
\caption{ Sequence of lobes between April 17,  2003 and May 26, 2003 mixing waters from north to south and viceversa at the Kuroshio current. }
\label{sequence}
\end{figure*}

Along the time interval under study, several lobes are formed mixing waters of  north and south.  Fig. \ref{sequence} contains a time sequence showing the evolution 
of several lobes created by the intersection of the stable and unstable manifolds. A sequence of trajectories $a,b,c, ...$ 
on the intersection points is depicted. These trajectories evolve from west to east and serve as reference to track lobe evolution. 
The lobes between the intersection trajectories $a$ and $b$  of  \ref{sequence}a) are not depicted in Fig.  \ref{sequence}b) as they are very thin filaments stretched along the 
DHT$_E$. In the same way the lobes between the 
intersection trajectories $c$ and $d$ of  Fig.  \ref{sequence}b) are not depicted in Fig. \ref{sequence}a) as they are very thin filaments stretched along the 
DHT$_W$. We notice that in  Fig.  \ref{sequence}d) the western hyperbolic trajectory is labelled as DHT$^+_W$ since  
 it is a different distinguished trajectory to DHT$_W$ that stopped as distinguished on May 11.  
Despite DHT$_W$ has lost  its distinguished property before we have completed our transport description,  the manifold computation
beyond this time  still makes sense as it
still is a material surface, and is still asymptotic to  DHT$_W$. 
Although  the manifold is not asymptotic to 
 DHT$^+_W$,  it is correct to say that DHT$^+_W$ marks a Distinguished Trajectory
 on the manifold  within certain accuracy $\epsilon$. Any further description on the turnstile
mechanism between May 11 and 26 needs to be referred to  this new DHT. 
 This kind of replacement of the DHT of reference along the invariant manifold has been discussed  in \cite{physrep}.  The piece of stable manifold
in  \ref{sequence}d) does not show more lobes on May 26, and beyond this day the turnstile mechanism is disrupted due to topological
transitions in the flow structure. 
 
The altimeter-derived estimates of the velocity field have expected errors,  so one question  is to establish whether  the information obtained from the analysis of manifolds is robust
to changes in the velocity field or not. The influence of deviations of the velocity fields on the Lagrangian structures
has been examined in several works \cite{haller3, ismael, pojel}. The main conclusion of these works is that Lagrangian structures are robust to errors, though
 conditions for the reliability of predictions are discussed. In what concerns to  lobes and its role on transport it is possible to say that these are
formed by intersections of stable and unstable manifolds and are typically observed  in the neighborhood of a Distinguished Hyperbolic Trajectory.  Despite the lack of  rigorous mathematical
theorems for aperiodic time dependent dynamical systems it is possible to foresee a relation between the vicinity of these DHTs  and 
  chaotic saddles.  These latter persist under small perturbations of the vector field, i.e. they are structurally stable \cite{wiggins}. For this reason 
  we also expect that the lobe description near the DHTs is
  robust under small  deviations on the measured velocity field.

\section{Conclusions}

We have explored the power of dynamical systems ideas for 
analyzing  altimeter data sets.  In particular  tools for quantifying Lagrangian transport have been successfully applied to   oceanographic observations.
Our study analyses cross-frontal transport in altimeter data sets along a flow located in the area of the Kuroshio current. 
  As a novelty with respect to similar studies we have  identified relevant DHTs following  the methodology 
proposed in a recent paper by \cite{chaos}, instead of using  that proposed in \cite{ide,nlpg,physrep,jpo}.  
 We have been able to detect relevant DHTs and  we confirm that
 DHTs hold the {\it distinguished} property only for  finite time intervals. 
 Identification and computation of 
DHTs  is by itself an
important subject, since they organize the flow in the area and,
because of this and of the sensitivity of the trajectories in
their neighborhood, they are candidates for launch locations in
efficiently designed drifter release experiments
\cite{ptkj,Molcard2006}.

We have computed  stable and unstable manifolds of the relevant DHTs. 
A time dependent Lagrangian  `barrier'  is defined from pieces of the stable and unstable manifolds and transport across 
it  is described by lobe
motion. It is found that the turnstile mechanism  is  at work in this observational flow, leading 
to a filamentous transport near the hyperbolic trajectories.  
This mechanism survives  between  April 17, 2003 and  May 26,  2003. 
After this date transitions in the flow take place, that are not well understood.
 
  In summary this article explores  the use, in a general oceanographic problem,   of
 ideas and techniques  associated with chaotic transport and dynamical systems, trying to shed some 
 light on the mechanisms that lead to messy particle trajectories over the ocean. We expect that this tools and 
 information  will be of interest  in the oceanographic community.

\section*{Acknowledgements}
C.M. and A.M.M are very grateful to Antonio Turiel for his valuable help with the altimeter data. 
Also we are indebted with his many insightful comments and suggestions.

The computational part of this work was done using the CESGA computers
SVGD and FINIS TERRAE.

The authors have been supported by CSIC Grant OCEANTECH No. PIF06-059, Consolider
I-MATH C3-0104, MICINN Grants Nos.   MTM2008-03754 and MTM2008-03840-E, and the Comunidad de Madrid
Project No. SIMUMAT S-0505-ESP-0158.

\addtocounter{figure}{-1}\renewcommand{\thefigure}{\arabic{figure}a}

\end{document}